\documentclass[aps,prb,reprint,amsfonts,superscriptaddress]{revtex4-2}
\usepackage{amsmath}  
\usepackage{bm} 
\usepackage{graphicx}
\usepackage{gensymb}
\usepackage{color,soul}
\usepackage{makecell}
\usepackage{multirow}
\usepackage{soul}
\usepackage{adjustbox}
\graphicspath{ {./images/} }

\usepackage[dvipsnames]{xcolor}

\newcommand{\vv}[1]{\bm{\mathrm{#1}}}

\begin{document}

\title{Spin-current mediated exchange coupling in MgO-based magnetic tunnel junctions}

\author{\L. Gladczuk}
\affiliation{Clarendon Laboratory, Department of Physics, University of Oxford, Oxford, OX1~3PU, United Kingdom}
\affiliation{Diamond Light Source, Harwell Science and Innovation Campus, Didcot, Oxfordshire OX11~0DE, United Kingdom}

\author{L. Gladczuk}
\affiliation{Institute of Physics, Polish Academy of Science, Aleja Lotnikow 32/46, PL-02668 Warsaw, Poland}

\author{P. Dluzewski}
\affiliation{Institute of Physics, Polish Academy of Science, Aleja Lotnikow 32/46, PL-02668 Warsaw, Poland}

\author{K. Lasek}
\affiliation{Institute of Physics, Polish Academy of Science, Aleja Lotnikow 32/46, PL-02668 Warsaw, Poland}

\author{P.~Aleshkevych}
\affiliation{Institute of Physics, Polish Academy of Science, Aleja Lotnikow 32/46, PL-02668 Warsaw, Poland}

\author{D. M. Burn}
\affiliation{Diamond Light Source, Harwell Science and Innovation Campus, Didcot, Oxfordshire OX11~0DE, United Kingdom}

\author{G.~\surname{van der Laan}}
\affiliation{Diamond Light Source, Harwell Science and Innovation Campus, Didcot, Oxfordshire OX11~0DE, United Kingdom}

\author{T. Hesjedal}
\affiliation{Clarendon Laboratory, Department of Physics, University of Oxford, Oxford, OX1~3PU, United Kingdom}

\date{\today}

\begin{abstract}
\noindent
Heterostructures composed of ferromagnetic layers that are mutually interacting through a nonmagnetic spacer are at the core of magnetic sensor and memory devices. In the present study, layer-resolved ferromagnetic resonance was used to investigate the coupling between the magnetic layers of a Co/MgO/Permalloy magnetic tunnel junction. 
Two  magnetic resonance peaks were observed for both magnetic layers, as probed at the Co and Ni $L_3$ x-ray absorption edges,  showing a strong interlayer interaction through the insulating MgO barrier.
A theoretical model based on the Landau-Lifshitz-Gilbert-Slonczewski equation was developed, including exchange coupling and spin pumping between the magnetic layers.
Fits to the experimental data were carried out, both with and without a spin pumping term, and the goodness of the fit was compared using a likelihood ratio test. 
This rigorous statistical approach provides an unambiguous proof of the existence of interlayer coupling mediated by spin pumping.
\end{abstract}

\maketitle

\section{Introduction}
Magnetic tunnel junctions (MTJs) are composed of two ferromagnetic films separated by an ultrathin layer of insulating material. Due to spin-dependent tunneling, the current flow in such systems is affected by the relative direction of the magnetization in the layers \cite{Butler2001, Tsymbal2003, Mathon2001}. MTJs have attracted much attention in the past years \cite{Moodera, Miyazaki1995, Parkin2004, Yuasa2004, Sato2012} as promising candidates for spintronic devices, offering a broad range of practical uses, e.g., at the core of sensor elements \cite{Naik2014} and memory devices \cite{Gallagher2006, Prejbeanu2013a}.
MgO-based MTJs are particularly attractive due to their high tunneling magnetoresistance ratio \cite{Butler2001, Yuasa2008, Chen2017}.

From a micromagnetic perspective, MTJs are characterized by the properties of the individual magnetic layers, as well as the interaction between them. Fundamentally, such an interlayer interaction can occur in three different ways: ({\it{i}}) through exchange coupling, i.e., the magnetic moments of neighboring magnetic layers directly interact with each other \cite{Stiles1999, Poulopoulos1999}; ({\it{ii}})  through spin-pumping, i.e., the spin current emitted by the precessing magnetization vector in one layer is driving the dynamics of the other layer; and ({\it{iii}})  through spin-transfer torque, i.e., an electrical current passing through the junction is spin polarized by the first layer and exerts torque on the second layer \cite{Slonczewski1996, Berger1996, Tsoi1998}.

%
Spin pumping, i.e., the flow of a pure spin current out of a ferromagnet undergoing ferromagnetic resonance, is an intriguing alternative to spin-transfer-torque driven devices as there is no net charge flow, and hence no Ohmic dissipation.
It can be observed via an increase in damping of the layer at resonance due to the outflow of angular momentum.
A pure spin current can persist across a normal metal or even an insulating barrier. 
In a trilayer structure, a spin current can be pumped from a source layer, passing through a non-magnetic spacer (normal metal or insulating layer), into a ferromagnetic sink layer whereby its strength decays exponentially with barrier thickness \cite{Du2013}.
To study spin pumping via the observation of antidamping torque in a trilayer structure, the magnetic resonances of the two layers have to be brought close together, making the fitting procedure in the data analysis challenging \cite{Baker2019}.

Spin pumping in common spin-based devices, such as MTJs, is of particular interest.
Previously, we found spin pumping in MTJs with a 1-nm-thick MgO barrier \cite{Baker2016SciRep}, contrary to earlier reports which concluded that MgO is a good spin sink \cite{Mosendz2010}, but confirmed later by electrical measurements on Fe/MgO/Pt \cite{Mihalceanu2017} and CoFeB/SiO$_2$/Pt \cite{Swindells2020}.
Owing to the strong damping in insulators, but motivated by the prospects of more energy efficient spin-transfer torque magnetic random access memory, a deeper understanding of the dynamic properties of MTJs has to be gained by refining the tools and methods used for their study.

Here, we shed new light on layer coupling and spin pumping phenomena by performing a detailed study of the magnetic layer dynamics in a Co/MgO/Permalloy (Py =  $\mathrm{Ni}_{80}\mathrm{Fe}_{20}$) MTJ heterostructure using layer- and time-resolved x-ray detected ferromagnetic resonance (XFMR) \cite{Arena2006, Marcham2013, Stenning2015, Li2016, Baker2016SciRep, Figueroa2016, Laan2017}.
XFMR employs x-ray magnetic circular dichroism (XMCD) to detect ferromagnetic resonance (FMR) from magnetic alloys and multilayers of thicknesses as thin as a few nanometers in an element-specific way \cite{Gerrit2014}.
For heterostructures composed of chemically distinct magnetic layers, each of the constituent layers can be probed separately.
XFMR is therefore ideally suited for characterizing interlayer coupling unambiguously in many device-relevant heterostructures.

The outline of the paper is as follows. In Sec.\ \ref{sec2:exp}, the sample preparation and the applied methods are discussed, followed by the theoretical model and the evaluation of the modeling parameters in Secs.\ \ref{sec3:model} and \ref{sec4:modpara}. Section \ref{sec5:XFMR} reports the XFMR results and the data are analyzed using a likelihood ratio test (Sec.\ \ref{sec6:robinhood}), followed by the conclusions in Sec.\ \ref{sec7:concl}.

\section{Sample preparation and experimental details\label{sec2:exp}}

\begin{figure}
	\centering
	\includegraphics[width = 0.48\textwidth]{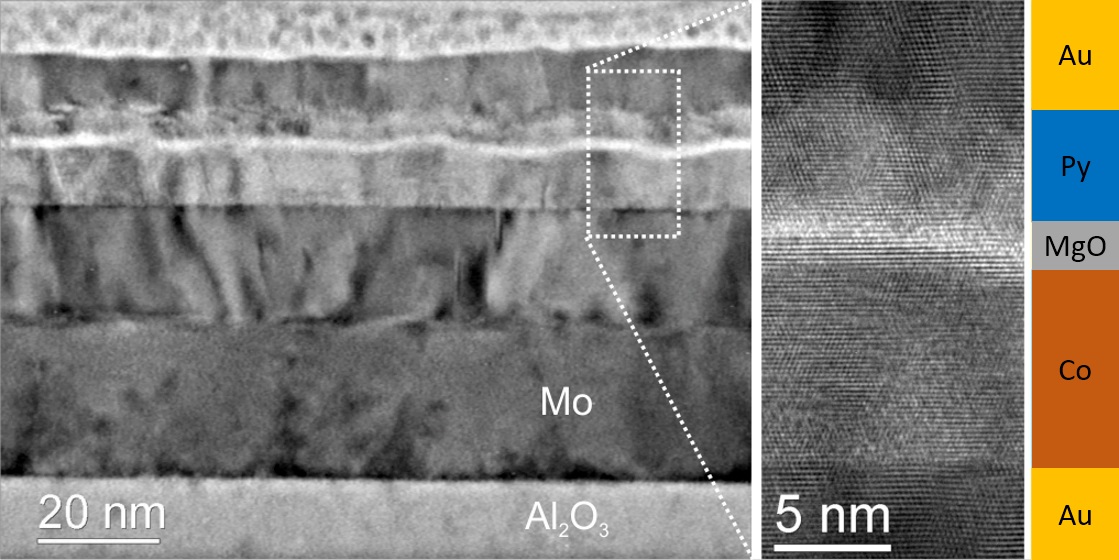}
	\caption{XTEM image of the $a$-plane sapphire/Mo/Au/Co/MgO/Py/Au MTJ. 
		In the overview on the left-hand side, the layer stack can be seen. In the close-up on the right-hand side, the core of the MTJ is highlighted, consisting of the ferromagnetic Co and Py layers, separated by a continuous, insulating MgO barrier.}
	\label{fig:TEM}
\end{figure}

Samples were grown using molecular beam epitaxy (Riber EVA 32, base pressure of $1 \times 10^{-10}\,$Torr). The base layers were prepared on polished \textit{a}-plane sapphire substrates (CrysTec). To ensure high quality films, a proven recipe was used \cite{Gladczuk2013, Gladczuk2014}: the substrate was covered with a {20$\,$nm} thick Mo layer, followed by {20$\,$nm} Au. 
The surface of such a stack is known to provide an ideal template for hcp \textit{c}-plane Co growth. 
The 10\,nm Co layer was covered with {2$\,$nm} MgO as the insulating barrier.
On top of the insulator, a 5$\,$nm thick magnetic Py layer was deposited. The structure was capped with Au to prevent oxidation. The growth quality was monitored using reflection high-energy electron diffraction (RHEED) and a quartz microbalance was used for thickness control.
The structural properties of the samples were determined using cross-sectional transmission electron microscopy (XTEM) (see Fig.\ \ref{fig:TEM}).
The magnetic properties of the samples were studied using resonance cavity FMR, vector network analyzer (VNA)-FMR, and XFMR.

The present study focuses on an MTJ sample with the following composition: /\!/Au/Co(10)/MgO(2)/Py(5)/Au (the numbers in parentheses denote thicknesses in nm, and /\!/ indicates the a-plane sapphire/Mo/ substrate).
For comparison, /\!/Au/Co(10)/MgO(2)/Au and /\!/Au/MgO(2)/Py(5)/Au samples were grown, which differ from the MTJ in that only either the magnetic  under- or overlayer is present.
These additional samples were used to characterize properties of the individual magnetic layers. 

Cavity FMR measurements were performed at room temperature (300\,K) and at 80\,K using an X-band Bruker EMX spectrometer ($f = 9.38\,$GHz). The samples were mounted on a quartz sample holder and placed in a resonance cavity. The FMR microwave absorption spectra were recorded for the external magnetic field in range of 0-1.6\,T, and magnetic field directions between in-plane ($\theta_H = 90\degree$) and out-of-plane ($\theta_H = 0\degree$).
The data was used to extract the magnetic resonance fields, allowing for a comparison with the theoretical model.

\begin{figure}[]
\centering
\includegraphics[trim={0 0 0 0},clip,width = 0.4\textwidth]{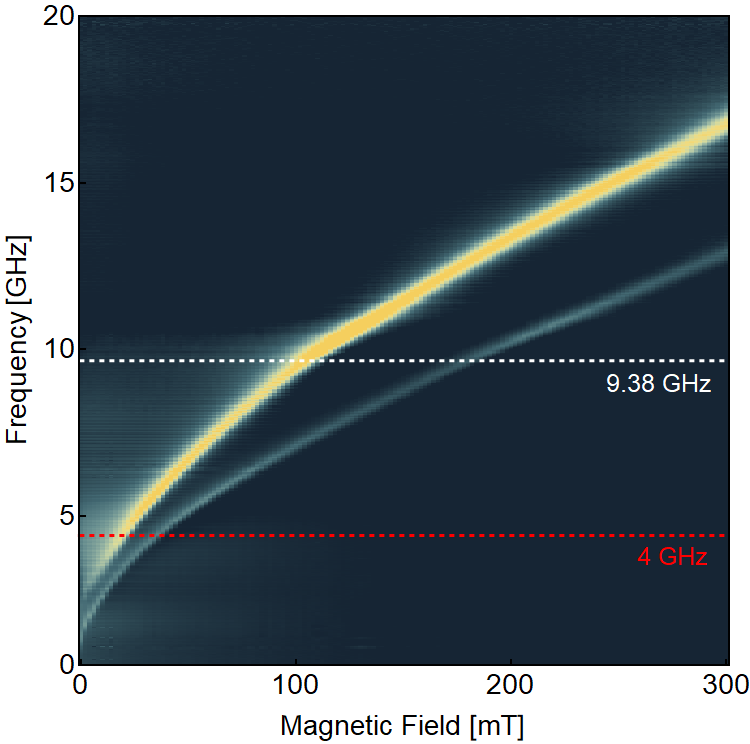}
\caption{
	VNA-FMR absorption as a function of applied magnetic field ($\theta_H = 45\degree$) and excitation frequency for the MTJ sample held at 300\,K. 
	The brightness of the signal corresponds to the strength of the absorption.
	As the sample consists of two distinct magnetic layers, two resonance curves are visible.
	White and red dashed lines indicate the frequencies at which the cavity FMR and XFMR were performed, respectively.
}
\label{fig:VNAFMR}
\end{figure}

\begin{figure}[]
	\centering
	\includegraphics[width = 0.45\textwidth]{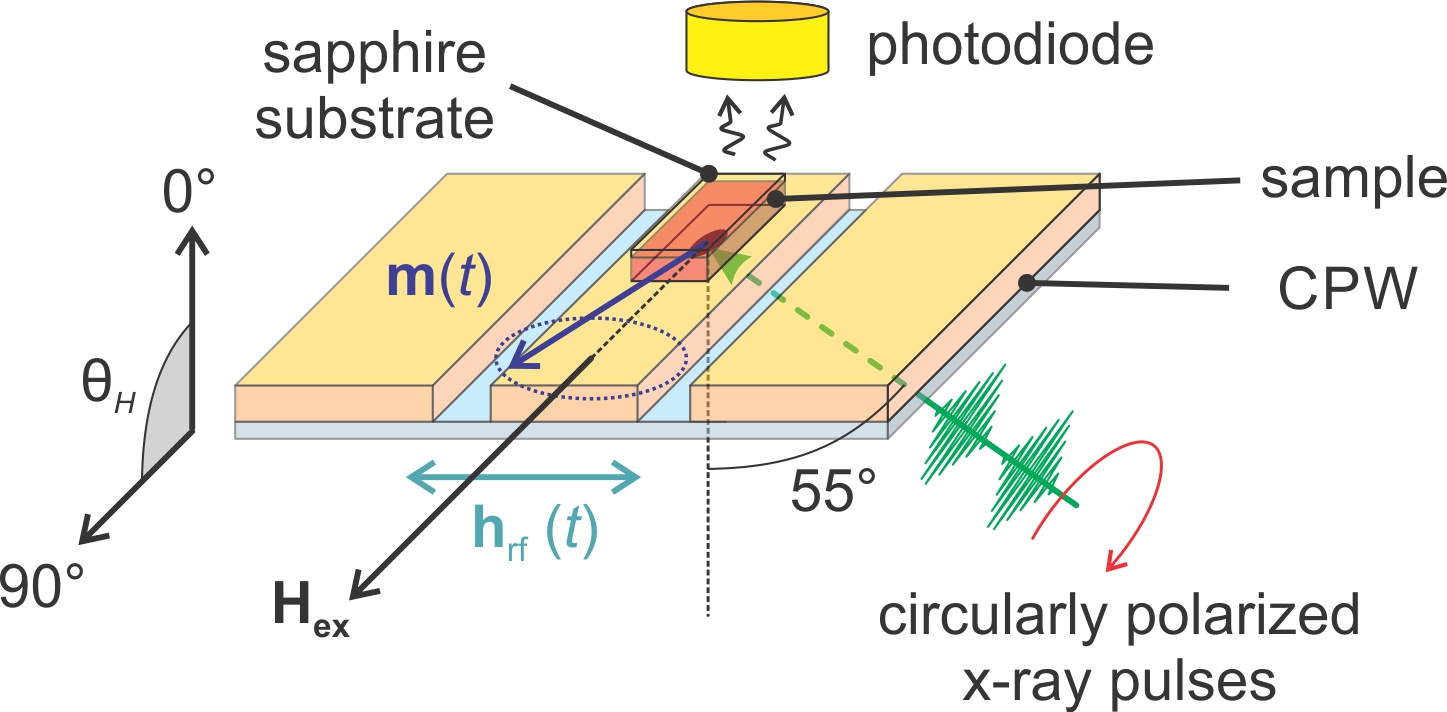}
	\caption{
		Schematic of the XFMR measurement in transverse  geometry.
		The sample (shown in red) is placed face down onto the central conductor of a coplanar waveguide (CPW), which has a tapered hole at the sample position.
		This allows circularly polarized x rays to enter onto the sample from below, under an angle of 55\degree\ with respect to the surface normal.
		A microwave signal $\vv{h}_{\mathrm{rf}}(t)$ is fed to the CPW and the magnetization $\vv{m}(t)$ precesses about the direction of the applied static magnetic field $\mathbf{H}_{\mathrm{ex}}$ (illustrated here for the $\theta_H = 90^\circ$ in-plane direction).
		Due to the pulsed nature of the synchrotron radiation, the component of $\vv{m}(t)$ along the beam direction $\vv{\hat{k}}$ can be detected stroboscopically via the measurement of the luminescence yield from the sapphire substrate using the photodiode behind the sample.
	}
	\label{fig:setup}
\end{figure}

VNA-FMR measurements were performed with the sample placed onto a holder with a coplanar waveguide (CPW), avoiding direct electric contact between the two. The holder was inserted into an octopole magnet capable of applying a magnetic field in any direction.
Using the VNA, a radio-frequency (rf) signal was applied to the CPW and the absorption spectra of the samples were measured in the range between 0.5-20\,GHz. 
The measurements were performed for bias fields ranging from 0 to 300$\,$mT (with a reference measurement performed at 400$\,$mT), and magnetic field directions ranging from in-plane to out-of-plane in steps of $5\degree$ (see Fig.\ \ref{fig:setup}), and at temperatures between 80 and 300$\,$K. 
From the data, the field-frequency maps of the rf absorption are extracted (Fig.~\ref{fig:VNAFMR}).

The x-ray detected FMR experiments were carried out on beamline I10 at the Diamond Light Source (Didcot, UK). 
In transmission geometry, the samples are placed face-down over a CPW containing a small tapered hole in the center conductor which allows the x rays to penetrate the waveguide and hit the sample, as illustrated in Fig.\ \ref{fig:setup}. 
The XFMR technique employs the XMCD effect for obtaining element-specific magnetic properties \cite{Gerrit2014}, by using  circularly polarized x rays   with the photon energy tuned to  the Co or Ni $L_3$ absorption edge. 
The x rays that are not absorbed by the MTJ, but instead transmitted, convert by x-ray excited optical luminescence  in the sapphire substrate into optical light, which is detected using a photodiode behind the sample. This signal provides a measure for the x-ray absorption in the magnetic layers, and using circular polarization probes the magnitude and direction of the magnetization vector. 
At the Diamond synchrotron, the x rays arrive in bunches at a rate of $\sim$500$\,$MHz, having an average pulse length of $\sim$34\,ps \cite{Thomas2006}. 
This time structure is used to perform stroboscopic measurements of the magnetic moment direction of the sample, thereby capturing its time-resolved dynamics. 
The driving frequency for the CPW needs to be an integer multiple of the synchrotron frequency, and the phase is referenced against the phase of the synchrotron master oscillator clock. The signal is filtered and amplified by performing lock-in detection, while modulating the phase of the driving magnetic field by 180$\degree$ at a frequency of $\sim$1\,kHz. 
All XFMR measurements presented in this work were performed at an rf driving frequency of 4$\,$GHz, with the normal of the sample surface oriented at an angle of $55\degree$ with respect to the beam direction. 

\section{Theoretical model for magnetic moment precession\label{sec3:model}}

The dynamics of the magnetic layers in the MTJ is modeled using the Landau-Lifshitz-Gilbert-Slonczewski equation, in which two types of interlayer coupling are considered, i.e., Heisenberg-type exchange coupling (EC), and coupling due to spin current (SC), i.e., spin pumping mediated coupling \cite{Tserkovnyak}
\begin{equation}
\small
\label{eq:Motion}
\begin{split}
\frac{\partial \vv{m}_i}{\partial t} = &-\gamma_i \,\vv{m}_i \times \left(\vv{H}_{\mathrm{eff},i} + \vv{h}_\mathrm{rf} + \frac{J_i}{\mu_0}\vv{m}_j \right)\\
&\quad+\alpha_i\,\vv{m}_i \times \frac{\partial \vv{m}_i}{\partial t}\\
&\quad\quad + a^{\uparrow\downarrow}_i\left(\vv{m}_i \times \frac{\partial \vv{m}_i}{\partial t} - \vv{m}_j \times \frac{\partial \vv{m}_j}{\partial t}\right)\;,
\end{split}
\end{equation}
where $\vv{m}_i = \vv{M}_i/M_i$ is the unit vector in the direction of magnetization of the $i^\mathrm{th}$ magnetic layer ($i$ = Co (Py), $j$ = Py (Co)), $\gamma_i$ is the electron gyromagnetic ratio and $\alpha_i$ is the layer-specific dimensionless coefficient of Gilbert damping.
 $\vv{H}_{\mathrm{eff},i}$ denotes the contribution of the external magnetic field $\vv{H}_\mathrm{ex}$,  the demagnetization field $\vv{H}_{\mathrm{dem},i}$, and  the anisotropy field $\vv{H}_{\mathrm{ani},i}$ of the $i^\mathrm{th}$ magnetic layer. 
Without an external source of energy acting on the system,  the stored potential energy will dissipate and the system will stabilize at an equilibrium position. 
To counteract the damping, a time-dependent magnetic driving field $\vv{h}_\mathrm{rf}$ is applied. The contribution from the Heisenberg-type coupling is parameterized using the EC parameter $J_i$, whereas the spin pumping contribution SP is described using the parameter $a^{\uparrow\downarrow}_i$ \cite{Tserkovnyak}.
The effect of both types of coupling is not the same between the layers and is scaled by the inverse product of the layer thickness $d_i$ and the saturation magnetization $M_i$.
This can be included in the model by introducing $\tan\zeta = \left(d_\mathrm{Py} M_\mathrm{Py}\right)/\left(d_\mathrm{Co} M_\mathrm{Co}\right)$, and defining: $J_\mathrm{Co} = J \csc\zeta$, $J_\mathrm{Py} = J \sec\zeta$, $a^{\uparrow\downarrow}_\mathrm{Co} = a^{\uparrow\downarrow} \csc\zeta$, and $a^{\uparrow\downarrow}_\mathrm{Py} = a^{\uparrow\downarrow} \sec\zeta$. In this way the precise dependence of the model on $d_i$ and $M_i$ is embodied in a single parameter $\zeta$.
It is assumed that the external magnetic field, layer magnetization, and other magnetic materials parameters are uniform within each layer.

The anisotropy of the magnetic Co and Py layers is assumed to be uniaxial with a single trigonometric expansion term in case of Py ($K^\mathrm{Py}_1$), and two trigonometric expansion terms in case of Co ($K^\mathrm{Co}_1$, $K^\mathrm{Co}_2$). 
Under these assumptions, the anisotropy energy (per unit volume) is given by
\begin{align}
E_\mathrm{ani,Py} &= -K^\mathrm{Py}_1 \left(\vv{m}_\mathrm{Py} \cdot \vv{n}\right)^2   \, , \\
E_\mathrm{ani,Co} &= -K^\mathrm{Co}_1 \left(\vv{m}_\mathrm{Co} \cdot \vv{n}\right)^2 - K^\mathrm{Co}_2 \left(\vv{m}_\mathrm{Co} \cdot \vv{n}\right)^4 \, , 
\end{align} 
where $\vv{n}$ denotes the unit vector normal to the sample surface. 
The demagnetization energy (per unit volume)  for each magnetic layer is assumed to have the form
\begin{equation}
E_{\mathrm{dem,}i} =  \frac{\mu_0 M_i^2}{2} \left(\vv{m}_i \cdot \vv{n}\right)^2 \,\,.
\end{equation}
The anisotropy (and demagnetization) field is related to the respective energy via
\begin{equation}
\vv{H}_{\mathrm{ani(dem),}i} = -\frac{1}{\mu_0 M_i}\frac{\partial E_{\mathrm{ani(dem),}i} }{\partial \vv{m}_i} \,\,\,\,.
\end{equation}

The dynamics of the system is found assuming that the magnetic driving field is weak ($h_\mathrm{rf}(t) \ll H_0$) and oriented in the sample plane, i.e., $\vv{h}_\mathrm{rf} = \hat{\vv{y}} \, h_\mathrm{rf} \exp(i\omega t)$. 
Equation \eqref{eq:Motion} is solved by linearization around the equilibrium direction $\vv{m}_{i,0}$, and by assuming periodic motion of the magnetization vectors $\vv{m}_i(t) = \vv{m}_{i,0} + \Re[\delta \vv{m}_i \exp(i\omega t)]$. 
The equilibrium of the magnetization direction is found by minimizing the potential energy of the system. 
In general, multiple local energy minima can exist.
Here, we chose the direction corresponding to the global energy minimum. 

The resulting amplitude and phase of the oscillation of the magnetization vector are compared to the experimentally determined quantities. In case of cavity FMR, the absorbed power $P_\mathrm{FMR}$ can be related to the dynamics of the magnetic layers using the relationship
\begin{equation} \label{P-FMR}
\small
\begin{split}
P_\mathrm{FMR} \propto 
&-\mu_0 M_\mathrm{Co} \frac{d_\mathrm{Co}}{d_\mathrm{Co}+d_\mathrm{Py}} \left<
\vv{m}_\mathrm{Co} \cdot \frac{d}{d t}\vv{H}_\mathrm{eff,Co} \right>\\
&\quad-\mu_0 M_\mathrm{Py} \frac{d_\mathrm{Py}}{d_\mathrm{Co}+d_\mathrm{Py}} \left<
\vv{m}_\mathrm{Py} \cdot \frac{d}{d t}\vv{H}_\mathrm{eff,Py} \right> \,\,,
\end{split}
\end{equation}
where the quantity inside the angle brackets is averaged with respect to time.
In the case of XFMR measurements, the projection of the precessing magnetization vector $\vv{m}_i$ onto the x-ray beam direction $\vv{\hat{k}}$ is proportional to
\begin{equation} \label{I-XFRM}
I_{\mathrm{XFMR},i}(t) \propto \vv{m}_i(t) \cdot \vv{\hat{k}} \,\,\,.
\end{equation}

The exact expressions for Eqs.~\eqref{P-FMR} and \eqref{I-XFRM} in terms of system parameters and magnetization direction are rather lengthy and will not be presented here. Instead, we refer the eager reader to the derivations in Ref.\ \cite{ThesisLG}.

\section{Estimation of model parameters\label{sec4:modpara}}

The  materials parameters introduced in Sec.\ \ref{sec3:model} can be determined through fitting to the experimental data. This usually amounts to a serious practical challenge, in which multiple complementary experimental techniques have to be used to determine all relevant parameters. 

Here, the thickness of the individual layers was determined using  XTEM. For the MTJ sample, the values are: $d_\mathrm{Co} = (9.2\pm1.2)\,\mathrm{nm}$, $d_\mathrm{Py} = (4.6\pm1.0)\,\mathrm{nm}$, and $d_\mathrm{MgO} = (1.4\pm0.8)\,\mathrm{nm}$, which are close to the nominal values of 10, 5, and $2\,\mathrm{nm}$, respectively.

The VNA-FMR measurements allow us to capture the magnetization dynamics simultaneously in the dependence on the external magnetic field and the frequency of $\vv{h}_{\mathrm{rf}}$.
The FMR of the MTJ sample shows two distinct magnetic resonance modes (Fig.\ \ref{fig:VNAFMR}). 
Since the layers are coupled, each of the layers contributes to the FMR signal at each mode.
The lower Kittel curve (i.e., the one with lower frequencies at any given field) corresponds to the acoustic mode, in which the magnetic moments of the two layers precess roughly in-phase. The upper Kittel curve corresponds to the optical mode, in which the magnetic moments of the layers precess roughly in antiphase \cite{Stenning2015}. 
The shape of these curves and their dependence on the external magnetic field direction ($\theta_H$), assuming uniaxial anisotropy, suggest that both magnetic layers are characterized by a negative anisotropy coefficient $K_1$, i.e., the in-plane direction of layer magnetization is energetically favored. This is in agreement with the expected magnetic properties of the Co and Py layers. 
Note that if the coupling is not too strong, each of the modes can be related to the intrinsic resonances of the individual layers (as if they were isolated).
Such an association is possible because the approximate values for the layer anisotropies are known from previous experiments.
We will confirm this  identification in Sec.~\ref{sec5:XFMR} by making use of the layer specificity of XFMR.


\begin{figure*}[htbp]
\centering
\includegraphics[width = 0.99\textwidth]{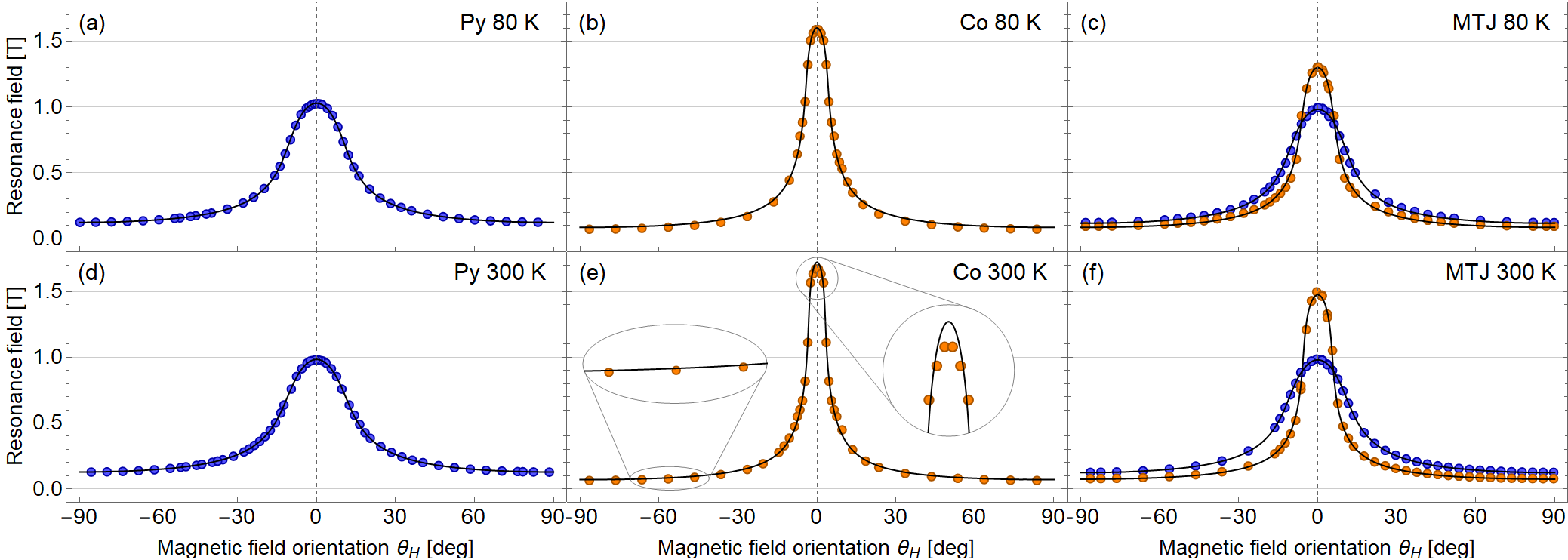}
\caption{Cavity FMR resonance curves as a function of external magnetic field direction ($\theta_H$) with respect to the surface normal for the Py, Co, and MTJ samples at 80 and  $300\,\mathrm{K}$.
	Blue and orange dots refer to measurement points for Py and Co, respectively. The black lines represent the theoretical model fits, yielding the fitting parameters summarized in Table \ref{tab:FMRTable}.}
\label{fig:FMR}
\end{figure*}

\begin{table*}
\caption{Anisotropy parameters $D_i$ and gyromagnetic ratios $\gamma_i$ obtained from fitting the theoretical model to the experimental cavity FMR data. 
Values for the Co/MgO/Py, Co/MgO, and MgO/Py samples are shown for measurements conducted at 300 and 80\,K. 
The theoretical relationship is expressed in terms of $D_1 = \left(2K_1+4K_2\right)/M$ and $D_2 = -4K_2/M$, which eliminates the need to provide the value of the layer magnetization.
In case of the MTJ sample, zero coupling between the two magnetic layers was assumed.
}
\begin{ruledtabular}
\def\arraystretch{1.2}
\centering
\small
\begin{tabular}{ccccccc}
 Sample & Fixed temp. & $\gamma_\mathrm{Py}$ & $D^\mathrm{Py}_1$ & $\gamma_\mathrm{Co}$ & $D^\mathrm{Co}_1$ & $D^\mathrm{Co}_2$ \\
 & [K] & [m\,s$^{-1}$A$^{-1}$] & [T] & [m\,s$^{-1}$A$^{-1}$] & [T] & [T] \\
 \hline
  Co/MgO/Py & 300 & $237\,040 \pm 230$ & $-0.668\pm0.001$ & $240\,950 \pm 920$ & $-1.207\pm0.016$ & $0.040\pm0.017$\\
 Co/MgO/Py & 80 & $243\,800 \pm 1\,000$ & $-0.678\pm0.005$ & $250\,400\pm 1\,400$ & $-0.942\pm0.017$ & $-0.061\pm0.019$\\
 Co/MgO & 300 & --- & --- & $255\,800\pm 2\,200$ & $-1.134\pm 0.039$ & $-0.298\pm 0.039$\\
 Co/MgO & 80 & --- & --- & $238\,600\pm 1\,600$ & $-1.052\pm 0.030$ & $-0.238\pm 0.032$\\
 MgO/Py & 300 & $232\,350 \pm 290$ & $-0.665\,\pm 0.002$ & --- & --- & ---\\
 MgO/Py & 80 & $231\,380\pm 170$ & $-0.709\pm 0.001$ & --- & --- & ---\\
\end{tabular}
\end{ruledtabular}
\label{tab:FMRTable}
\end{table*}

The available field range of the VNA-FMR data did not permit a precise determination of the anisotropy parameters and gyromagnetic ratio.
Instead, cavity FMR was employed for their determination, as this setup allowed for higher magnetic fields.
%
%
%
%
The absorbed rf power is recorded as a function of magnetic field direction $\theta_H$ and strength $H$, and the resonance magnetic field at given $\theta_H$ is determined.
In principle, Eq.\ \eqref{eq:Motion} can be used to derive the relationship between the resonance field and $\theta_H$. 
However, fitting such a relationship to the experimental MTJ data is challenging since the fitted anisotropy constants are dependent on the chosen value of the effective gyromagnetic ratio. 
The latter is sensitive to interfacial effects \cite{Sasage_2010} and can incorporate interlayer coupling.
In order to obtain a useful estimate for the effective gyromagnetic ratios, the complementary single magnetic layer reference samples /\!/Au/Co(10)/MgO(2)/Au and /\!/Au/MgO(2)/Py(5)/Au were studied and analyzed.
For convenience, normalized anisotropy values are introduced as $D_1 = \left(2K_1+4K_2\right)/M$ and $D_2 = -4K_2/M$.
The parameters $D_i$ and $\gamma_i$ were determined using the conjugate gradient method (as implemented in Mathematica 12.1) to fit the experimental magnetic resonance fields (Fig.\ \ref{fig:FMR}, Table \ref{tab:FMRTable}).

For the MgO/Py sample, a good fit of the experimental data was obtained assuming uniaxial anisotropy. 
The fitted values for the gyromagnetic ratio and anisotropy constant are very similar for 300 and 80\,K, and such a temperature independence is characteristic for high-quality Py.
In case of the Co/MgO sample, however, the fit was slightly less good for low and high values of $\theta_H$. This can be best seen for the 300\,K data in the inserts to Fig.\ \ref{fig:FMR}(e), where the fitted value for low and high angles is larger than the experimental data, and smaller than the experimental data in range of $10\degree<|\theta_H|<30\degree$. 
The resulting gyromagnetic ratio for the Co sample at $300\,\mathrm{K}$ is larger than reasonable.
This might be due to the fact that the Co layer could be a mixture of the hcp and fcc phases. In this case, the Co anisotropy will no longer be uniaxial.
Although it seems to be a minor issue at this point, a higher precision is required for the XFMR analysis in Sec.~\ref{sec5:XFMR}.

Next, the theoretical dependence on the resonance field as a function of external magnetic field direction was fitted to the obtained data for the two layers.
Unfortunately, it is not possible to extract with sufficient confidence values for $J$ and $a^{\uparrow\downarrow}$.
We have therefore  in the FMR fitting set these values equal to zero, i.e., assuming a zero-coupling scenario in which the layer dynamics is independent.
The resulting parameter values are listed in Table \ref{tab:FMRTable}.


The values of the gyromagnetic ratio and the anisotropy constants found for the MTJ are different from those obtained for the single magnetic layer samples. This can be caused partly by a small difference in layer thickness between the MTJ and the Co and Py samples.
The values of the gyromagnetic ratio for the Co layer are high, especially in case of the MTJ sample at $80\,\mathrm{K}$.
This could be an indication that coupling is present in the system.
As the model used for the fitting does not explicitly include coupling, their effects become incorporated into the other fitting parameters, such as the gyromagnetic ratio.
Thus for an in-depth study of the layer coupling and a more reliable extraction of the model parameters, more detailed experimental data is needed, which XFMR is able to provide.


\section{XFMR Measurements \label{sec5:XFMR}}

The Gilbert damping parameter $\alpha_i$ and the magnetic layer coupling parameters $J$ and $a^{\uparrow\downarrow}$ can be obtained from the XFMR data.
First, in an XFMR experiment, delay scans are performed, i.e., the XFMR signal is measured as a function of phase difference (delay) between the incoming x-ray pulses and the phase of the driving signal fed to the CPW (and, in turn, the phase of precessing magnetization vector).
Since the strength of the XFMR signal in transverse geometry depends on the projection of the magnetic moment onto the beam direction, such measurements allow for capturing the time-resolved magnetization dynamics.

\begin{figure*}[htbp]
\centering
 \includegraphics[width = 0.99\textwidth]{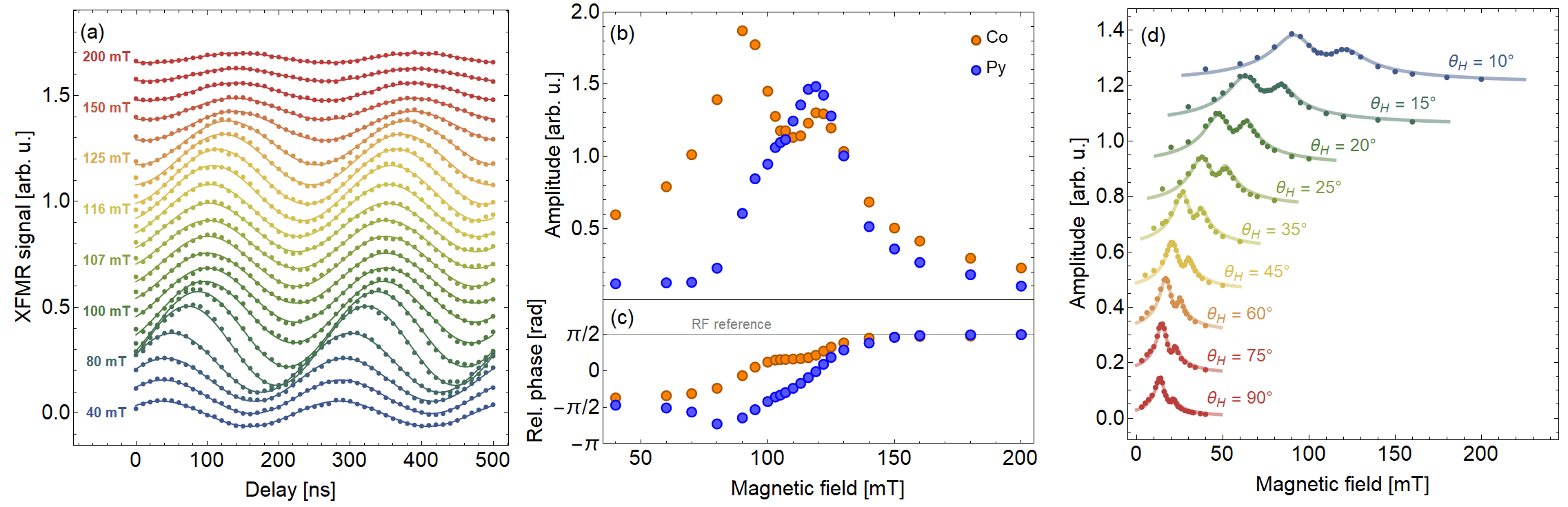}
%
\caption{XFMR measurements of the MTJ sample at 80\,K. 
	(a) XFMR delay scans for the Co layer with the magnetic field oriented at $\theta_H = 10\degree$.
	For clarity, the data points (circles) obtained at different magnetic field values are shifted by a constant offset. 
	The continuous lines represent the fitted sinusoidal functions.
	Their amplitude and phase as a function of magnetic field strength is plotted in panels (b) and (c), respectively, for both the Co (orange) and Py (blue) layers.
	(d) Magnetic field direction dependence of the XFMR amplitude for the Co layer. The in-plane measurement at $\theta_H = 90\degree$ corresponds to panel (b).}
\label{fig:XFMR}
\end{figure*}

Delay scans were performed for each of the magnetic layers at fixed magnitude and direction of the magnetic field. A sinusoidal function of the form
\begin{equation} \label{sinfit}
S(t) = X \sin(2\pi f t) + Y \cos(2\pi f t) \, ,
\end{equation}
was fitted to the data, where $t$ represents the time delay and $f$ is the frequency of excitation (here 4\,GHz).
This procedure was repeated for various field strengths and directions. By extracting the coordinates $X$ and $Y$ in Eq.\ \eqref{sinfit} from the delay scans, the amplitude and phase of the oscillations can be determined as a function of $H$ and $\theta_H$, using the relationships
\begin{equation} \label{eq:Iphase}
\small
C = \sqrt{X^2+Y^2},\ \ \ \ \ \psi=2\arctan\left(\frac{Y}{\sqrt{X^2+Y^2}+X}\right) \,\,.
\end{equation}
Some experimental data of the Co layer precession in MTJ sample are shown in Fig.\ \ref{fig:XFMR}(a).

The choice of the form for the function $S(t)$ in Eq.~(\ref{sinfit}) is not incidental. The sine and cosine functions are orthogonal, thus the estimators of $X$ and $Y$ are given by projections to orthogonal subspaces. This results in the uncertainty for the estimation of $X$ and $Y$ to be independent, meaning that the mean squared error (MSE) is a simple sum of the squared residuals in $X$ and $Y$.
Additionally, the standard error of the regression ($\sigma$) together with $X$ and $Y$ are also independent. 
Although it may seem more natural to fit a function of the form $C\sin(2\pi f t +\psi)$, this leads to nonlinear estimators for amplitude $C$, phase $\psi$, and error $\sigma$. Consequently, the noise values associated with the estimation of $\psi$ and $C$ are interrelated.

The sinusoidal fits match the measured XFMR delay scan data well, allowing for the extraction of the amplitude and phase of the Co and Py layer oscillations [Fig.\ \ref{fig:XFMR}(b) and \ref{fig:XFMR}(c)]. 
Each of the amplitude curves reveals two peaks, where the larger one corresponds to the intrinsic FMR mode of the layer. The presence of a smaller peak at the position of the main peak in the other layer is evidence for the interaction between them.
The coupling effect is observed for all field angles, with a $\theta_H$-dependent shift in the position and   width of the resonance peak, which are due to the anisotropy of the system [Fig.\ \ref{fig:XFMR}(d)].

The phases plotted in Fig.\ \ref{fig:XFMR}(c) reveal an overall phase shift of $+\pi$ across the full resonance spectrum.
The detailed structure of the phase should in principle show a sigmoidal-shaped feature in case of spin pumping and a bell-shaped dip in case of exchange coupling \cite{Marcham2013, Li2016}.
Here, however, the two modes are too close together and are overlapping, so that such a straightforward and simple interpretation of the data is not on offer. 
%
On the high field side of the resonance spectrum, the Py and Co signals are in phase with the rf field, serving as a phase reference value. 
Lowering the field to $\sim$120\,mT, the resonance of the acoustic mode is reached where the Py amplitude signal is strong in magnitude [Fig.\ \ref{fig:XFMR}(b)] and $\sim$$\pi /2$ out-of-phase with the driving rf signal.
The Co layer is in-phase with the Py precession up to the Py signal maximum, at which point the Co phase is lagging $\pi /2$ behind the rf, and its amplitude reaches a local minimum. 
Above the Py resonance, i.e., on the low field side, the Py signal is in antiphase with the rf field, i.e., its phase is shifted by $\sim$$\pi$. 
The Co precession is then no longer in phase with that of the Py layer, and remains almost constant at $\sim$0\degree.
With decreasing field, the Co phase smoothly approaches $-\pi/2$, while the Py phase shows a further drop close to $-\pi$, before recovering again and joining the Co phase at very low fields.

\begin{figure*}[htbp]
	\centering
	\includegraphics[width = 0.99\textwidth]{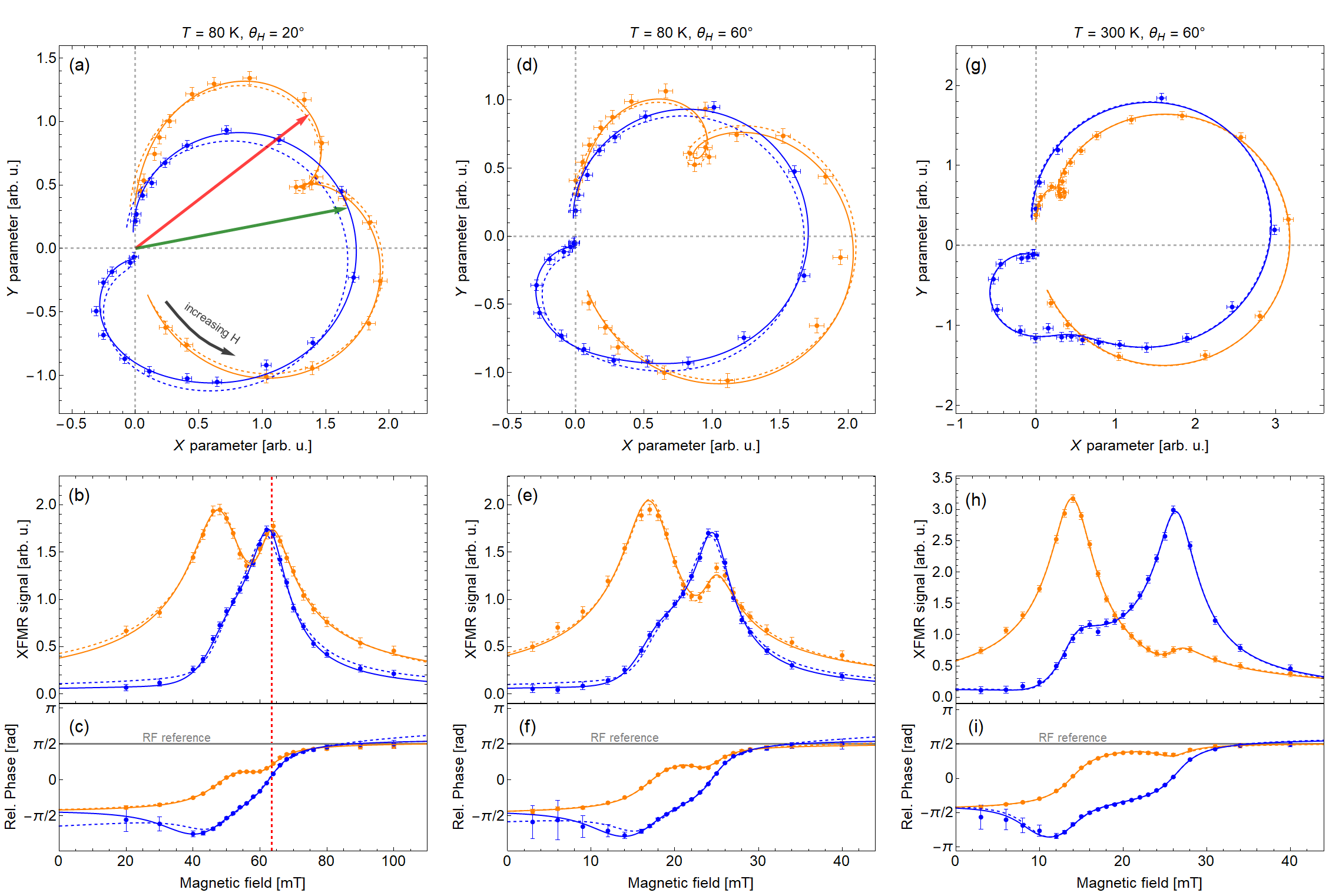}
	\caption{Comparison of the results for model I (EC only, dashed lines) and model II (EC+SC, solid lines) for different external magnetic field angles and temperatures: (a-c) $\theta_H = 20\degree$, $T= 80$\,K; (d-f) $\theta_H = 60\degree$, $T= 80$\,K; and (g-i) $\theta_H = 60\degree$, $T= 300$\,K.
	The experimental data is shown as circles for the Co (orange) and Py (blue) layers.
	The top row shows the FMR plots in the $(X,Y)$-plane, for the field increasing going counterclockwise along the curves.
	The middle and bottom row show the corresponding amplitude ($C$) and relative phase ($\psi$) plots, respectively, as a function of field.
The red and green vectors in panel (a) relate to the $X$ and $Y$ values in model II for Co and Py, respectively, at a magnetic field of $63.5\,$mT. The length of the vectors corresponds to the signal amplitude and the angle with the horizontal axis corresponds to its phase. 
For comparison, the same value of magnetic field is indicated by red dashed lines in panels (b) and (c).
}
\label{fig:TeoVExp}
\end{figure*}

The theoretical model introduced in Eq.\ \eqref{eq:Motion} can be directly used to determine the coefficients $X$ and $Y$ in the sinusoidal function $S(t)$ of Eq.\ \eqref{sinfit}.
%
The XFMR signal in the $(X,Y)$-plane, as shown in Figs.\ \ref{fig:TeoVExp}(a,d,g), resembles a Cayley's sextic curve, whereby each of the semicircles corresponds to a different mode of precession.
The amplitude of the signal ($C$) can be found as the distance of a point from the origin of the plot, and the phase ($\psi$) as the angular position of the point with respect to the horizontal axis [see Eq.\ \eqref{eq:Iphase}].
The magnetic field, which  increases along the circles in the $(X,Y)$-plane in the counterclockwise direction, is hard to visualize in the plots.
Instead, for clarity, we show the amplitude  ($C$) and phase ($\psi$) plots in Fig.\ \ref{fig:TeoVExp} underneath each $(X,Y)$-plot as a function of magnetic field.

For the fitting procedure, the measurement uncertainty (i.e., the standard error of the regression, $\sigma$) is assumed to be identical for the $X$ and $Y$ coordinates, and it is fitted independently for each of the magnetic layers (Co and Py).
%
To account for the absolute scale of the XFMR signal ($C_{\theta_H}^i$), the relative phase of the oscillations ($\psi_{\theta_H}^i$), and a sinusoidal background noise ($r_{\theta_H}^i$, $q_{\theta_H}^i$), additional model parameters are introduced as
%
\begin{equation} \label{equ:XFMR_MODEL}
\small
\begin{split}
X^i_{\theta_H}(H) = r_{\theta_H}^i + \Re\left[C_{\theta_H}^i \delta \vv{m}_i \cdot \vv{\hat{k}} \exp\left(i\psi_{\theta_H}^i\right)\right] \, , \\
Y^i_{\theta_H}(H) = q_{\theta_H}^i + \Im\left[C_{\theta_H}^i\delta \vv{m}_i \cdot \vv{\hat{k}} \exp\left(i\psi_{\theta_H}^i\right)\right] \, ,
\end{split}
\end{equation}
where $i$ denotes the Co or  Py layer.
The quality of the fit is very sensitive to the values of the anisotropy constants, and even small adjustments can produce a much better fit and a more accurate estimation of the coupling parameters.
Note also that the cavity FMR data already hinted that the Co layer is not a system with uniaxial anisotropy.
These issues are resolved by assuming that the anisotropy of the magnetic layers is instead given by
\begin{equation}
\begin{split}
E_\mathrm{ani,Py} = -K_{\theta_H}^\mathrm{Py}\left(\vv{m}_\mathrm{Py} \cdot \vv{n}\right)^2 \, , \\
E_\mathrm{ani,Co} = -K_{\theta_H}^\mathrm{Co}\left(\vv{m}_\mathrm{Co} \cdot \vv{n}\right)^2  \, ,
\end{split}
\end{equation}
where the parameters $K_{\theta_H}^i$ were fitted independently for each $\theta_H$ angle. Similarly as before, normalized anisotropy constants, $D_{\theta_H} = 2K_{\theta_H}/M$, are introduced.

\begin{table*}
\caption{Fit parameters for the MTJ sample measured at 300 and 80\,K. For each temperature, the upper and lower row show the values for model I (EC only) and model II (EC and SP), respectively.
$\tan\zeta = \left(d_\mathrm{Py} M_\mathrm{Py}\right)/\left(d_\mathrm{Co} M_\mathrm{Co}\right)$.
}
\begin{ruledtabular}
\def\arraystretch{1.2}
\begin{tabular}{ccccccccccc}
Temp. &  
Model &
\makecell{$\alpha_\mathrm{Co}$ $[10^{-3}]$} & 
\makecell{$\alpha_\mathrm{Py}$ $[10^{-3}]$} & 
\makecell{$\tan\zeta$} & 
\makecell{$J$ $[10^{-3}\,\mathrm{kg}\,\mathrm{A}^{-1}\,\mathrm{s}^{-2}]$} &
\makecell{$a^{\uparrow\downarrow}$ $[10^{-3}]$} \\\hline
\multirow{2}{*}{300\,K} 
& EC only & $15.22\pm0.14$ & $16.59\pm0.13$ & $0.129 \pm 0.007$ & $0.615\pm 0.023$ & --- \\
& EC+SP & $12.26\pm0.66$ & $15.65\pm0.24$ & $0.130 \pm 0.007$ & $0.620\pm 0.022$  & $0.411 \pm 	0.090$ \\\hline
\multirow{2}{*}{80$\,$K}
& EC only & $14.50\pm0.20$ & $20.42\pm0.29$ & $0.269 \pm 0.021$ & $0.861\pm 0.022$ & --- \\
& EC+SP & $6.26\pm0.58$ & $17.65\pm0.30$ & $0.126 \pm 0.008$ & $0.668\pm 0.020$ & $1.115 \pm 	0.053$
\end{tabular}
\end{ruledtabular}
\label{tab:XFMRfit}
\end{table*}

Two models were investigated to verify the existence and importance of spin pumping in the MTJ. Model I only includes EC coupling, i.e., the SP parameter is set to zero, while model II includes both EC and SP coupling. 
The theoretical dependencies [Eq.\ \eqref{equ:XFMR_MODEL}] were fitted directly to the $X$ and $Y$ values, defined in Eq.\ \eqref{sinfit}, using the conjugate gradient method with respect to the following parameters: $J$, $a^{\uparrow\downarrow}$, $\zeta$, $\alpha_i$, $D_{\theta_H}^i$, $C^i$, $\psi^i$, $r^i$, $q^i$, and $\sigma^i$. 
The values for $\gamma_i$ at 80 and 300\,K were determined by cavity FMR to be $\gamma_\mathrm{Py}(80\,\mathrm{K})$ = 231\,380, $\gamma_\mathrm{Py}(300\,\mathrm{K})$ = 232\,350, $\gamma_\mathrm{Co}(80\,\mathrm{K})$ = 238\,600, and $\gamma_\mathrm{Co}(300\,\mathrm{K})$ = 240\,000\,m\,s$^{-1}$A$^{-1}$.
Since discrepancies in the $\gamma_\mathrm{Co}$ values were noticed, a sensible arbitrary value was chosen. The values of the anisotropy constants found from cavity FMR were used as initial conditions for the fitting procedure.

In order to explore the influence of $\gamma_\mathrm{Co}$ on the determination of the other parameters further, the fitting procedure was repeated for $\gamma_\mathrm{Co}$ in the range from 228\,000 to 250\,000\,m\,s$^{-1}$A$^{-1}$ (in steps of 400\,m\,s$^{-1}$A$^{-1}$).
It was found that the variation of $\gamma_\mathrm{Co}$ affects the fitted anisotropy constants $D_{\theta_H}^i$, however, $J$ and $a^{\uparrow\downarrow}$ remain unaffected.
The fit parameters are summarized in Table \ref{tab:XFMRfit}, and the results for both models together with the experimental data are visualized in Fig.\ \ref{fig:TeoVExp}.

	The XFMR resonance data presented in Fig.\ \ref{fig:TeoVExp}, comparing resonance spectra for $\theta_H=20$\degree\ at 80\,K and $60$\degree at 80 and 300\,K, respectively, also allow for a study of the influence of magnetic field direction and temperature on the resonances.

	The effect of the field orientation on the XFMR signal can be seen by comparing Figs.\ \ref{fig:TeoVExp}(a,b,c) and  \ref{fig:TeoVExp}(d,e,f) (both obtained at 80\,K).
	Most importantly, as already illustrated in Fig.\ \ref{fig:XFMR}(d), the position of the resonances shifts continuously with applied field angle, ranging from $\sim$100\,mT at  $\theta_H=10$\degree\ (almost out-of-plane) to $\sim$15\,mT at $\theta_H=90$\degree\ (in-plane field).
	For the values shown in Fig.\ \ref{fig:TeoVExp}, the acoustic mode resonance occurs at $\sim$65\,mT for $\theta_H=20\,$\degree  [Figs.\ \ref{fig:TeoVExp}(b,c)], which reduces to $\sim$25\,mT for $\theta_H=60$\degree. 
	The amplitude of the optical mode (observed in the Co layer) is almost independent of the magnetic field direction, whereas the acoustic mode gets considerably stronger for fields applied more towards the surface normal direction.

	The effect of temperature can be seen by comparing the 80\,K data in Figs.\ \ref{fig:TeoVExp}(d,e,f) with the 300\,K data in Figs.\ \ref{fig:TeoVExp}(g,h,i)], which were both obtained at a field angle of 60\degree.
	Most strikingly, the relative strength of the acoustic mode (observed in the Co layer) as compared to the optical mode is much reduced at 300\,K, i.e., the relative effect of coupling between the layers appears to be weaker at higher temperatures. 
	Furthermore, at 300\,K, there is almost no difference between the fit results obtained with models I and II, i.e., spin pumping plays a much lesser role at higher temperatures.
	This finding is also supported by comparing the fitted $a^{\uparrow\downarrow}$ values of model II (shown in Table \ref{tab:XFMRfit}). For 300\,K, $a^{\uparrow\downarrow}$ is more than twice smaller than the 80\,K value.

When dealing with two different types of coupling in the MTJ, it is useful to provide a value that can quantify their relative importance for the system dynamics. From the equations of motion given in Eq.\ \eqref{eq:Motion}, it is directly found that the EC and SP coupling parameters can be grouped together in the form of
\begin{equation} \label{eq:couplig}
J \gamma_i g(\theta_{\mathrm{Co}},\theta_{\mathrm{Py}}) + i \mu_0 \omega a^{\uparrow\downarrow} \, ,
\end{equation}
where $\cos\theta_i = \vv{m}_i \cdot \vv{n}$ and 
$g(\theta_{\mathrm{Co}},\theta_{\mathrm{Py}})$ are terms depending on the magnetization orientation of the layers ($g$ has the form of the quotient of trigonometric functions of $\theta_\mathrm{Co}$ and $\theta_\mathrm{Py}$). 

As $\vv{m}_i$ changes depending on the value of the external magnetic field, so do the contributions of EC ($J$) and SP ($a^{\uparrow\downarrow}$) to the coupling between the magnetic layers. With this in mind, it would be practical to provide a measure for the relative importance of EC and SP that depends only on the sample properties and could be used to compare different magnetic systems (i.e., independent of $\vv{H}_\mathrm{ex}$). We propose a dimensionless quantity $L_g$, defined below, that can be obtained from  Eq.\ \eqref{eq:couplig} when the $g$ factors are discarded, and that allows to estimate the relative importance of the EC and SP,
\begin{equation}
L_g = \frac{\mu_0}{\gamma}\frac{\omega a^{\uparrow\downarrow}}{J} \, ,
\end{equation}
where $\gamma = 221\,200\,$m\,s$^{-1}$A$^{-1}$ is the gyromagnetic ratio of the free electron. It is evident that the importance of the SP with respect to the EC is dependent on the oscillation frequency $\omega$. At low frequencies, EC will always dominate ($L_g\ll1$), whereas SP can become the major coupling mechanism at high frequencies ($L_g\gg1$). 

Plugging in the values found for the MTJ (in Table \ref{tab:XFMRfit}, EC+SP model), it is found that $L_g$(80\,K) = 0.23 and $L_g$(300\,K) = 0.09, which suggests that SP contributes to roughly 18.5\% and $8.3\%$ of the coupling, respectively.

\section{Likelihood ratio test\label{sec6:robinhood}}

The validity of competing theoretical models is evaluated by their ability to describe the observed data more accurately. This is not always an easy task, especially when several models appear to describe the data well.
In such a case, the goodness of fit can be quantified using statistical likelihood-ratio tests \cite{koch1999parameter}.

For the investigation of spin pumping in an MTJ, two competing models were evaluated. Model I assumes that only EC is present in the system, i.e., $a^{\uparrow\downarrow} \equiv 0$, whereas model II assumes that both EC and SP coupling are present.
In general, a model containing more fit parameters will always provide a better fit to the data, thereby achieving a better goodness of fit.
However, even if their inclusion is physically justified, it needs to be carefully evaluated if the improvement is significant enough to obtain a proof of their existence, i.e., spin pumping and $a^{\uparrow\downarrow} \neq 0$ in the present case. 

Such a dilemma can be resolved by preforming a likelihood ratio (LR) test, which allows for the quantification of the improvement, thus allowing to decide between the validity of both models
\begin{equation}
\lambda_\mathrm{LR} = -2\ln\left[\frac{\sup_{\vv{\xi}_0 \in \Xi_0}\mathcal{L}(\vv{\xi}_0)}{\sup_{\vv{\xi} \in \Xi}\mathcal{L}(\vv{\xi})}\right] \,\,.
\end{equation}
The numerator and denominator are the maximal likelihoods that can be achieved by fitting each model to the observed data.

In a statistical framework, where measurement points are considered to be random variables, $\lambda_\mathrm{LR}$ is also a random variable (whose probability distribution will be denoted as $\Lambda$). Here, the \textit{null hypothesis} parameter space $\Xi_0$ of model I is a subset of the \textit{alternative hypothesis} parameter space $\Xi$ of model II, and both models differ by a single degree of freedom. Therefore, as stated by Wilks' theorem, as the sample size approaches infinity, $\Lambda$ asymptotically approaches the $\chi_1^2$ distribution of the null hypothesis \cite{Wilks_1938}.

\begin{figure}[]
\centering
\includegraphics[trim={0 0 0 0},clip,width = 0.48\textwidth]{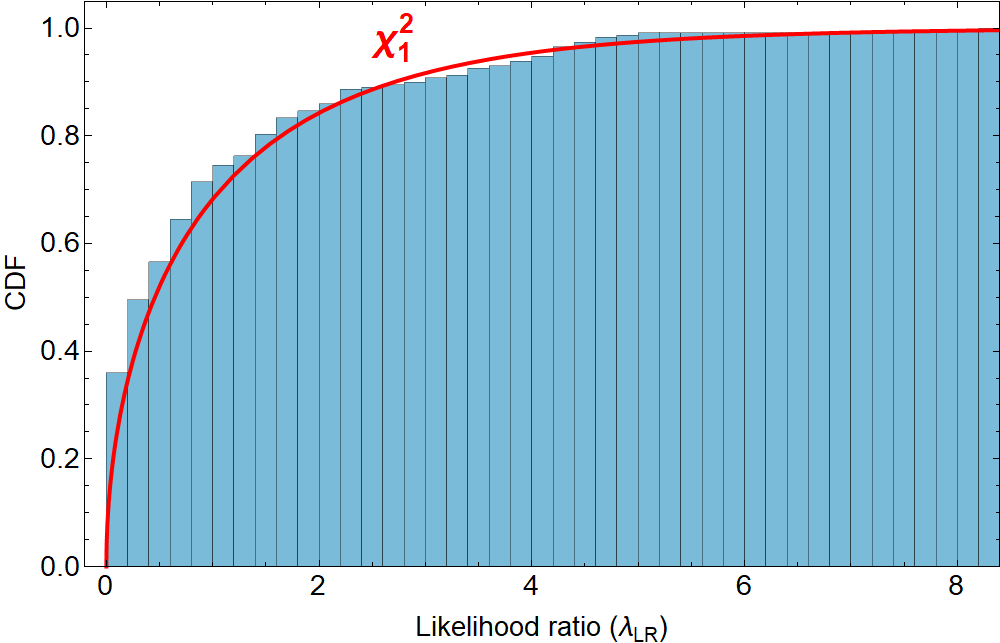}
\caption{Cumulative distribution function (CDF) of the simulated likelihood ratio ($\lambda_{LR}$), plotted together with the CFD of $\chi^2_1$.}
\label{fig:LR}
\end{figure}

To verify the results for a finite sample size (finite number of magnetic fields at which the XFMR signal is measured), we test the validity of the approximation of $\Lambda$ by the $\chi_1^2$ distribution.
First, assuming the \textit{null hypothesis} is correct, an artificial data set is created and both models are fitted to the created data set.
The resulting LR value for this artificial set is recorded and the procedure is repeated 230 times.
This results in a sample dataset from the $\Lambda$ distribution (Fig.\ \ref{fig:LR}). This distribution of the dataset is then analyzed and compared to the $\chi_1^2$ distribution using the Kolmogorov-Smirnov test \cite{kendall1994k}, resulting in a $\emph{p}$-value of 0.87.
This does not allow us to reject, at a $5\%$ level, the hypothesis that the sample data is distributed according to the $\chi^2_1$ distribution, thus this confirms that such an approximation is valid.

Having approximated $\Lambda$ with $\chi_1^2$, it is possible to determine the confidence by which the \textit{null hypothesis} can be rejected.
This is done by evaluating the cumulative distribution of $\Lambda$ at the value of $\lambda_\mathrm{LR}$ obtained in the real experiment. For the measured MTJ sample, the resulting LR values were: $\lambda_\mathrm{LR}^{80\,\mathrm{K}} = 575.7$ and $\lambda_\mathrm{LR}^{300\,\mathrm{K}} = 77.6$, with the corresponding $\emph{p}$-values of $4.0\times10^{-127}$ and $1.4\times10^{-18}$, respectively. These $\emph{p}$-values allow us to state with great confidence that the \textit{null hypothesis} can be rejected, i.e., requiring model II in which both types of coupling are presents to describe the experimental MTJ data.

\section{Conclusions\label{sec7:concl}}

An in-depth study of Co/MgO/Py MTJs was performed using layer- and time-resolved XFMR.
A theoretical model, based on micromagnetic theory, was derived for the system under investigation. By fitting the model to the acquired data, parameters characterizing the magnetization dynamics of the MTJ were determined. 

The spin pumping mediated coupling between the magnetic layers has been confirmed by performing a rigorous statistical analysis.
Two alternative hypotheses were considered, the first only considering interlayer exchange coupling and the second also including spin pumping. The hypotheses were compared using a likelihood ratio test, allowing us to conclude with great confidence that spin pumping is present in the MTJs.
The improvement of the fit that is achieved by including spin pumping is most clearly visible at $80\,$K. The fitted spin pumping coefficient $a^{\uparrow\downarrow}$ is more than twice as high for the MTJ sample at $80\,$K than at $300\,$K. This shows that spin pumping is more effective at lower temperatures, which agrees with the theoretical understanding.
In summary, using a rigorous statistical approach for the analysis of XFMR data, we are able to provide unambiguous proof of the existence of spin pumping in MTJs, thereby supplying a reliable tool set for the systematic temperature-dependent study of spin transfer phenomena in heterostructures.\\

\section*{Acknowledgments}

We acknowledge the Diamond Light Source for beamtime on I10 under proposal numbers SI-21616 and SI-18759, and the Research Complex at Harwell for their hospitality. Ł.G.\ acknowledges funding from the Diamond Light Source through a joint studentship and the Engineering and Physical Sciences Research Council (EPSRC) through a doctoral training award. Ł.G.\ thanks R. Waszkiewicz for helpful discussions regarding the statistical methods used.

%

\end{document}